\documentclass[letterpaper]{article}

\PassOptionsToPackage{hyphens}{url}
\usepackage{flairs}
\usepackage{times}
\usepackage{helvet}
\usepackage{courier}
\usepackage{graphicx}
\usepackage{xcolor}
\begin{document}

\title{From Tokens to Ties: Network and Discourse Analysis\\of Web3 Ecosystems}
\author{Valentina Kuskova, Dmitry Zaytsev\\
Lucy Family Institute for Data \& Society\\
University of Notre Dame\\
Notre Dame, IN, 46556\\
}
\maketitle

\begin{abstract}
\begin{quote}
This paper examines Web3 ecosystems not merely as markets for digital assets, but as networked social spaces where economic transactions give rise to enduring social ties, shared narratives, and collective identities. Leveraging large-scale data mining of fused on-chain blockchain transactions and off-chain social media activity, we analyze over one hundred NFT collections to uncover how different forms of participation structure community formation in decentralized environments. Using network analysis, we identify distinct ecosystem roles, such as long-term holders, active traders, and short-term speculators, and demonstrate how each produces markedly different network topologies, levels of cohesion, and pathways for influence. We complement this structural analysis with discourse analysis of social media engagement, revealing how narrative production, visibility, and sustained interaction persist even as transactional activity declines. Our findings show that communities centered on holding behavior evolve from transactional networks into socially embedded ecosystems characterized by dense ties, decentralized influence, and ongoing cultural participation, while trader- and speculator-dominated networks remain fragmented and transactional. By linking network structure with discursive dynamics, this study provides a sociotechnical framework for understanding how value, identity, and inequality are negotiated in Web3 spaces. The approach offers a scalable method for detecting patterns of inclusion, exclusion, and representational imbalance, advancing network-based research on digital communities beyond purely economic or technical accounts.
\end{quote}
\end{abstract}

\section{Introduction}

Non-fungible tokens (NFTs) emerged into public awareness following several highly publicized transactions, including early record-setting sales of digital artworks (Kaczynski and Kominers, 2021). While these events initially framed NFTs as speculative digital assets, subsequent developments have revealed a more complex ecosystem in which economic exchange, social interaction, and cultural production are tightly intertwined. Rather than functioning solely as isolated transactions, NFT activity increasingly unfolds within persistent, networked communities that extend across blockchain platforms and social media.

Prior research has largely examined NFTs through a technological or application-oriented lens, emphasizing use cases such as copyright protection (\c{C}a\u{g}layan Aksoy and \"{O}zkan \"{U}ner, 2021), intellectual property commercialization (Okonkwo, 2021), healthcare systems (Musamih et al., 2022), game development (Fowler and Pirker, 2021), marketing (Chohan and Paschen, 2023), and fashion (Huggard and S\"{a}rm\"{a}kari, 2023). While this work highlights the versatility of NFT technology, it typically treats NFTs as discrete technical artifacts or economic instruments, paying limited attention to the social structures and interaction patterns that emerge around them at scale.

The rapid rise and subsequent market contraction of NFTs further underscores the need for such analysis. After an initial speculative boom, NFT markets experienced sharp declines in asset valuations, leading to widespread narratives of financial loss and market failure (Smith, 2023; Hategan, 2024). However, despite declining prices, NFT production and participation have continued to grow. Estimates suggest that the number of NFT collections increased substantially between 2023 and 2024, even as average transaction values fell (Minaev, 2023). This divergence between economic valuation and participation indicates that NFTs are being used for purposes beyond short-term speculation, motivating closer examination of their social and organizational dynamics.

Importantly, NFT ecosystems are not limited to on-chain transactions. They are accompanied by rich off-chain activity, including community discussion, narrative formation, and identity signaling across social media platforms. NFT collections often function as focal points around which participants coordinate, communicate, and establish shared meanings that persist independently of ongoing trading activity. These dynamics suggest that NFT ecosystems can be understood as sociotechnical networks, where value is produced not only through exchange, but also through social embedding and sustained interaction.

In this paper, we address this gap by analyzing NFT ecosystems as networked social systems rather than solely as markets. Leveraging large-scale data mining of fused on-chain transaction records and off-chain social media discourse, we examine over one hundred NFT collections to identify participation roles, network structures, and discursive dynamics. By combining network analysis with discourse analysis, we provide a framework for understanding how decentralized digital communities form, persist, and differentiate, offering insight into the emergence of value, identity, and inequality in Web3 environments.

\section{Socio-Technical Framing of NFT Ecosystems}

\begin{figure}[h!]
  \centering
  \includegraphics[width=\linewidth]{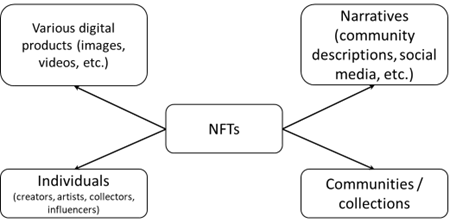}
  \caption{The many facets of NFTs.}
  \label{fig:nft}
\end{figure}

Prior research suggests that NFTs encompass more than digital assets or technical artifacts. As illustrated in Figure~\ref{fig:nft}, NFT ecosystems consist of multiple interrelated layers, including creators and buyers, on-chain transaction networks, off-chain communities, and ongoing narrative and discursive activity across social platforms. Together, these elements form integrated socio-technical systems in which technical infrastructure, economic exchange, and social interaction are mutually constitutive.

Beyond their role as transferable digital objects, NFTs have been used for cultural preservation and representation, including the digitization of historical artifacts (Metablock, 2024a), religious objects (Constantino, 2022), and Indigenous art (Crypto Altruism, 2023). Following the contraction of speculative NFT markets, such applications have gained increased prominence, suggesting a shift from purely transactional use toward community-oriented and meaning-driven engagement. These developments motivate analytical approaches that move beyond market behavior to examine how participation, identity, and interaction are structured within NFT ecosystems.

Digital art remains the most visible and widely adopted NFT application. Low production costs, global distribution, and liquidity enabled by blockchain infrastructure, often combined with creator anonymity, have been described as lowering barriers to artistic participation (Watkins, 2021) and, in some accounts, contributing to the democratization of art markets (Gibson, 2021). While such claims are contested, they underscore the extent to which NFTs intertwine creative production with networked participation and collective visibility.

Despite the rapid growth of NFT-related scholarship, work explicitly addressing their human-centered or sociocultural implications remains limited. A systematic review of Scopus-indexed publications in late 2022 identified only one Arts and Humanities article among 127 NFT-related studies (Taherdoost, 2022), with the remainder concentrated in computer science, engineering, economics, and social sciences. Where human-centered considerations do appear, they are typically operationalized through narrow fairness metrics, such as demographic representation among NFT holders (Young, 2020; Kumar, 2025), rather than through analysis of participation dynamics, discourse, or community structure.

More broadly, existing work rarely examines NFTs through alternative analytical lenses that emphasize recognition, participation, or communicative power, including identitarian (Fraser and Honneth, 2003), capability-based (Sen, 2001), discursive (Habermas, 1990), or meritocratic (Sandel, 2020) frameworks. This gap persists despite increasing recognition that AI-enabled and blockchain-based systems function as socio-technical systems that actively shape social relations rather than remaining neutral infrastructures (Dignum, 2022).

In this paper, we build on this perspective by analyzing NFT collections as networked socio-technical ecosystems. Rather than focusing on individual assets or price dynamics, we examine how participation roles, interaction patterns, and discursive practices co-evolve within NFT communities. By integrating network analysis with social media discourse analysis, we investigate how value, identity, and influence are produced and negotiated across decentralized environments, positioning NFTs as empirically tractable sites for studying community formation and inequality in Web3 systems.

\section{NFT Ecosystems as Community-Forming Technologies}

NFT-native properties, such as verifiable ownership, programmable access rights, and persistent on-chain identity, have enabled the rapid emergence of NFT-centered online communities. Rather than existing as isolated digital artifacts, NFT collections often serve as focal points around which participants organize social interaction, coordinate events, and develop shared narratives (Kaczynski and Kominers, 2021). These ecosystems extend beyond trading activity to include conferences, collaborative projects, and platform-specific social spaces, forming durable interaction networks that persist independently of asset price dynamics.

Despite their scale and visibility, NFT communities remain underexamined from a systems perspective. Unlike the traditional art world, where institutions and social roles are relatively transparent, NFT ecosystems operate largely through pseudonymous participation and decentralized coordination. This limits external visibility while simultaneously producing rich, traceable interaction data across blockchain and social platforms. As immersive online environments and metaverse-like systems continue to expand (Belk et al., 2022), understanding how such communities form, persist, and stratify has become an important empirical problem rather than a purely cultural one.

\section{Participation, Representation, and Opportunity Structures}

NFT ecosystems are frequently described as offering new opportunities for participation by historically underrepresented groups, due to lower barriers to entry, global reach, and reduced reliance on traditional gatekeeping institutions (Bastian, 2021). From a socio-technical perspective, this raises empirical questions about who participates in NFT ecosystems, how visibility and influence are distributed, and whether new forms of inequality emerge alongside new opportunities.

Existing discussions of inclusion in NFTs are largely anecdotal or project-specific, focusing on individual collections or platforms associated with Indigenous creators, Black artists, or women-led initiatives. While these cases illustrate the potential for alternative participation pathways, they do not provide systematic evidence about ecosystem-wide patterns of representation, community formation, or influence. Moreover, participation alone does not guarantee visibility, valuation, or sustained engagement, underscoring the need to distinguish between access, centrality, and long-term network position.

Rather than evaluating inclusion through categorical presence alone, this study treats participation as a measurable structural property of NFT ecosystems. By analyzing interaction networks and discourse dynamics, we examine whether communities associated with underrepresented creators occupy central or peripheral positions, whether influence is concentrated or distributed, and how narratives propagate across collections. This reframing shifts attention from individual success stories to population-level patterns that can be empirically assessed.

\section{Ethical and Equity Considerations in NFT Ecosystems}

Alongside their participatory potential, NFT ecosystems raise well-documented ethical and equity concerns. These include the circulation of harmful or exclusionary content, fraud and consumer protection issues, and environmental externalities associated with blockchain infrastructure (Ingold, 2022; Dean, 2022; Flick, 2022). From a sociotechnical standpoint, such issues are not peripheral but structurally embedded, reflecting how technical affordances interact with social norms and incentives.

Empirical research has begun to examine inequality in NFT markets, including evidence of racial and gender disparities in pricing and representation (Nguyen, 2022; Zhang et al., 2022). However, much of this work focuses on early or iconic collections, such as CryptoPunks, which may not be representative of the current NFT ecosystem. Given the rapid expansion and diversification of NFT collections, it remains unclear whether previously identified patterns generalize to newer communities or whether alternative structures have emerged.

More broadly, existing analyses of fairness and bias in AI-driven and blockchain-enabled systems often rely on narrow technical definitions that overlook contextual and discursive dimensions of inequality (Blodgett et al., 2020). This limits the ability to detect how exclusion, marginalization, or unequal amplification arise through participation dynamics rather than explicit design choices. In response, this study adopts an interdisciplinary, data-driven approach that integrates network structure and discourse analysis to examine how ethical and equity-related outcomes emerge within NFT ecosystems at scale.

\section{Data and Methodology for NFT Community Detection}

\subsection{Data Sources and Scope}

Data for this study were provided by Web3Sense (2025), a private analytics firm specializing in the integration of on-chain blockchain data with off-chain Web2 and open-source intelligence, including social media activity. The dataset covers all recorded transactions for hundreds of NFT collections, from the date of first mint through the date of data retrieval.

Each transaction record includes the unique token identifier, buyer and seller wallet addresses, transaction timestamp, transaction value denominated in Ethereum (ETH), gas fees, and, when available, Ethereum Name Service (ENS) identifiers. Although wallet addresses are pseudonymous by default, ENS registrations provide a partial linkage between on-chain behavior and publicly visible online identities, enabling richer contextual analysis.

\subsection{Network Construction and Community Detection}

Using transaction-level data, we constructed directed transaction networks in which nodes represent wallet addresses and edges represent NFT transfers between buyers and sellers. Edge weights capture transaction frequency and value, allowing differentiation between sporadic and sustained interactions. Specifically, edge weight is computed as a composite of transfer count and cumulative ETH value between a given wallet pair, enabling the network to reflect both activity intensity and economic significance simultaneously.

Graph layouts for all visualizations were generated using force-directed algorithms, specifically the ForceAtlas2 layout as implemented in Gephi, which positions nodes based on attraction along weighted edges and repulsion between all node pairs. This approach causes highly interconnected wallets to cluster together visually and isolated nodes to drift to the periphery, making community structure directly legible in the spatial arrangement. Consequently, visual differences across the holder, trader, and speculator network figures reflect genuine structural differences in interaction patterns, not arbitrary drawing choices.

To identify structurally coherent subgroups, we applied community detection techniques including modularity-based clustering (Louvain method) and label propagation. These methods reveal clusters of wallets that interact more frequently with one another than with the broader network, signaling shared ownership patterns, coordinated behavior, or collective engagement with specific collections. Such clusters serve as the basis for identifying distinct community types within NFT ecosystems. The numerical labels visible in Figure~\ref{fig:nft5} correspond to community cluster identifiers assigned by the detection algorithm, providing a basis for cross-community comparison of size, density, and centralization.

\subsection{Integration of On-Chain and Off-Chain Signals}

To contextualize transactional behavior, we enriched the on-chain networks with off-chain metadata using Web3Sense's identity-matching layer. This layer links wallet addresses to public social media accounts through ENS identifiers, user-declared wallet tags, and triangulated digital footprints.

This integration enables joint analysis of behavioral signals, such as minting frequency, resale patterns, and holding duration, with discursive indicators including posting volume, engagement levels, and participation in community spaces such as Twitter Spaces or Discord events. The dual-layer approach allows us to examine how identity signaling, social participation, and economic behavior co-evolve within NFT communities, rather than treating transactions as isolated events.

\section{Behavioral Typologies in NFT Ecosystems}

Analysis across dozens of NFT transaction networks revealed three recurring behavioral community types: holders, traders, and speculators. These types are not predefined categories but empirically emergent patterns derived from network structure, transaction dynamics, and temporal behavior.

\subsection{Holder Communities}

\begin{figure}[!htbp]
  \centering
  \includegraphics[
    width=\linewidth,
    height=0.35\textheight,
    keepaspectratio
  ]{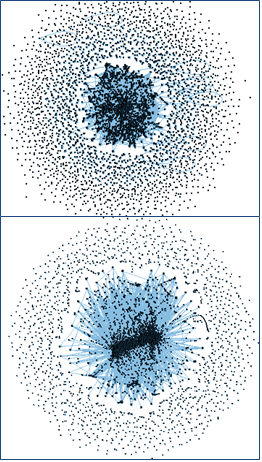}
  \caption{Communities of ``Holders'' -- Veefriends (top) and ArtBlocks (bottom).}
  \label{fig:nft2}
\end{figure}

Holder communities (Figure~\ref{fig:nft2}) consist of wallets that acquire NFTs and retain them over extended periods, exhibiting limited resale activity. These networks are characterized by dense cores of interconnected wallets surrounded by peripheral nodes that have purchased tokens but remain transactionally inactive thereafter.

The structure suggests more than transactional engagement; it reflects social cohesion and the emergence of digital affinity groups. These holders may be bound by shared aesthetic preferences, collective narratives, or community values, forming the cultural backbone of the NFT ecosystem. The network's shape, consisting of dense center with radiating tendrils, illustrates a vibrant and resilient social fabric, with the potential for influence, governance, and storytelling emerging from within. This is not merely a collection of users but a digitally-native community bound by belief in the symbolic and cultural value of the assets they hold.

\subsection{Trader Communities}

\begin{figure}[!htbp]
  \centering
  \includegraphics[
    width=\linewidth,
    height=0.35\textheight,
    keepaspectratio
  ]{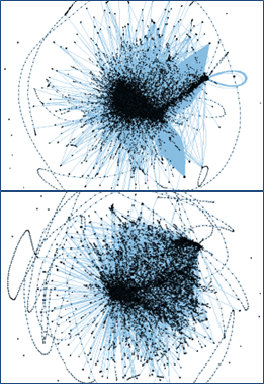}
  \caption{Communities of ``Traders'' -- Doodles (top) and CoolCats (bottom).}
  \label{fig:nft3}
\end{figure}

In contrast, trader communities (Figure~\ref{fig:nft3}) display highly dynamic and interconnected structures driven by frequent buying and selling across multiple collections. These networks feature dense transactional cores, extensive cross-cutting ties, and long transactional chains indicative of liquidity-driven behavior.

The presence of loops, bridges, and extended paths suggests arbitrage strategies, coordinated trading, or automated activity. Unlike holder communities, traders exhibit weak attachment to specific collections or narratives, operating instead as connectors across disparate ecosystems. Structurally, these networks function as high-throughput conduits for value transfer rather than as sites of sustained social cohesion.

\subsection{Speculator Communities}

\begin{figure}[!htbp]
  \centering
  \includegraphics[
    width=\linewidth,
    height=0.35\textheight,
    keepaspectratio
  ]{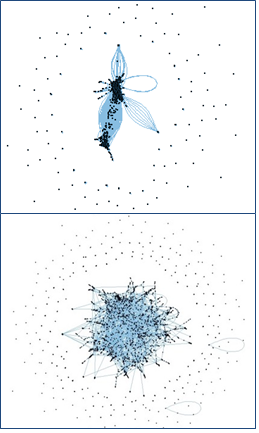}
  \caption{Communities of ``Speculators'' -- Finiliar (top) and BudHeritage (bottom).}
  \label{fig:nft4}
\end{figure}

Speculator communities (Figure~\ref{fig:nft4}) occupy an intermediate but distinct position. Characterized by short holding periods and episodic engagement, these networks show looser clustering, fragmented cores, and a high prevalence of transient nodes.

The structure reflects opportunistic participation driven by hype cycles, influencer signals, or trending indicators. Unlike traders, speculators exhibit limited strategic coordination, and unlike holders, they display minimal long-term attachment to community norms or narratives. Their fragmented topology and detached subgroups suggest low commitment and weak social integration, contributing to volatility and trust erosion within broader NFT ecosystems.

\section{Comparative Cluster Structure and Network Centralization}

\begin{figure}[!htbp]
  \centering
  \includegraphics[
    width=\linewidth,
    height=0.35\textheight,
    keepaspectratio
  ]{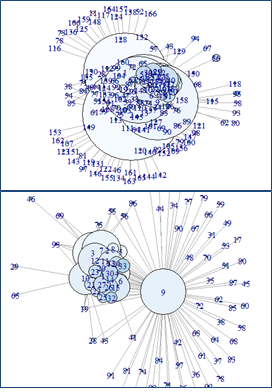}
  \caption{Clusters of Artblocks (top) and Doodles (bottom).}
  \label{fig:nft5}
\end{figure}

Cluster-level analysis further highlights structural differences across community types (Figure~\ref{fig:nft5}). The Art Blocks holder network exhibits a decentralized, interwoven cluster structure with multiple moderately sized hubs, indicating distributed influence and high mutual connectivity. This pattern is consistent with participatory cultures where engagement and authority are broadly shared.

In contrast, the Doodles trader network displays a hub-and-spoke configuration dominated by a single central node, with many peripheral wallets connected primarily through that hub. Such centralization facilitates efficient trade but limits redundancy, mutual visibility, and durable community ties. These contrasts illustrate how different participation modes generate fundamentally different social architectures within NFT ecosystems.

Together, these contrasting structures reinforce the central argument: holder communities like Art Blocks foster decentralized, richly connected networks conducive to cultural and social development, while trader communities such as Doodles exhibit hub-and-spoke models that facilitate liquidity but lack the structural complexity necessary for sustained community building. This methodology can be extended to assess whether vulnerable or underrepresented groups are participating primarily at the center or periphery of these networks, offering a new lens to evaluate not just inclusion, but integration.

\section{Temporal Dynamics of On-Chain and Off-Chain Activity}

\begin{figure}[!htbp]
  \centering
  \includegraphics[
    width=\linewidth,
    height=0.35\textheight,
    keepaspectratio
  ]{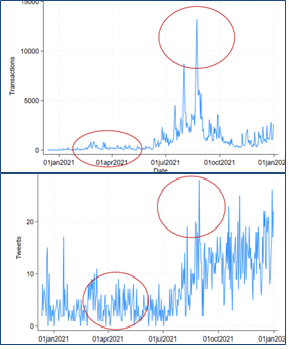}
  \caption{ArtBlocks community activity on-chain (top) and on X (formerly Twitter) (bottom).}
  \label{fig:nft6}
\end{figure}

To examine the relationship between economic and social engagement, we analyzed temporal patterns of transaction volume and social media activity. Figure~\ref{fig:nft6} juxtaposes on-chain transaction counts and Twitter activity for the Art Blocks community throughout 2021.

Both series show initial surges following minting and during peak market attention. However, while transaction volume declines after mid-2021, social media activity continues to rise and remains elevated. This decoupling indicates a transition from speculative engagement to sustained community participation, where discourse, interpretation, and identity-building persist independently of trading intensity.

This pattern is consistently observed across multiple holder communities but is largely absent in trader- and speculator-dominated networks, where social engagement typically tracks transaction volume more closely. The divergence highlights the role of holding behavior in catalyzing durable cultural ecosystems rather than transient market activity.

In the case of Art Blocks, social media activity became the scaffolding for a vibrant, durable holder network rooted in shared aesthetic appreciation and cultural participation. These individuals remained invested not just in the asset, but in the cultural and social life of the project, sharing interpretations of the artwork, organizing virtual events, and reinforcing the community's collective identity. The divergence between economic and discursive behavior in this data reinforces the idea that NFT communities like Art Blocks thrive not only through sales, but through storytelling, shared values, and lasting social bonds.

\section{Discussion}

Our analysis demonstrates that integrating network analysis with off-chain discourse data provides a robust sociotechnical framework for examining bias, inequality, and representation in NFT ecosystems. This approach enables simultaneous investigation of structural participation patterns and discursive visibility - dimensions that are often treated separately or overlooked in technically oriented fairness analyses.

From a structural standpoint, transaction networks reveal how influence and access are distributed across NFT communities. Network metrics such as centrality, modularity, and connectivity make it possible to identify whether participation is broadly distributed or concentrated among a small number of actors, as well as whether certain groups occupy peripheral or weakly connected positions. The presence of isolated clusters or limited cross-community ties may indicate systemic segmentation, even in ecosystems that appear open at the level of entry.

Off-chain discourse analysis complements these findings by capturing how communities are represented and engaged with in public-facing spaces. Patterns in language use, sentiment, and engagement volume reveal whose narratives gain traction and whose remain marginal. By linking transactional roles to social visibility through ENS identifiers and platform metadata, the framework distinguishes between nominal participation and meaningful influence in shaping community narratives.

Taken together, this socio-technical approach moves beyond surface-level demographic counts to examine how inequality emerges through interaction structure, discursive amplification, and network position. It provides a scalable and empirically grounded method for assessing fairness and inclusion in decentralized digital ecosystems, where social outcomes are produced through the interaction of technical infrastructure and collective practice.

\section{Conclusion and Research Implications}

This study demonstrates that Web3 ecosystems, and NFT collections in particular, can be systematically analyzed as sociotechnical networks in which economic transactions give rise to durable social structures and shared narratives. By mining fused on-chain transaction data and off-chain social media activity, we show that different participation modes - holding, trading, and speculation - produce distinct network topologies, levels of cohesion, and influence dynamics.

Our results indicate that holder-centered ecosystems transition from transactional interaction graphs into socially embedded networks characterized by dense ties, decentralized influence, and sustained discourse, even as trading activity declines. In contrast, trader- and speculator-dominated ecosystems remain structurally fragmented and temporally volatile, with limited persistence of social ties or narrative continuity. These findings highlight the importance of integrating behavioral roles and discourse signals into network-based analyses of decentralized digital communities.

Methodologically, this work contributes a scalable framework for linking network structure with discursive dynamics in Web3 environments. By combining role-based network analysis with social media discourse mining, the approach enables the detection of inclusion, exclusion, and representational imbalance as emergent properties of interaction rather than as purely economic outcomes. This moves beyond asset-centric or transaction-only analyses and provides a richer basis for studying value formation, identity signaling, and inequality in decentralized systems.

Future work will extend this framework longitudinally and across platforms, enabling causal and temporal analysis of community evolution, narrative diffusion, and structural change. Among the promising directions is the application of graph neural networks (GNNs), for example, modeling transaction networks with wallets, collections, and social accounts as heterogeneous node types, with transaction, mention, and ownership relations as edge types, which could support tasks such as community role prediction and link prediction across Web3 ecosystems. More broadly, the proposed approach offers a generalizable methodology for studying how social organization and cultural participation emerge from large-scale interaction data in decentralized online ecosystems.

\section*{References}

\smallskip \noindent
Bastian, R. 2021. The Diversity, Equity And Inclusion Potential Of NFTs. \url{https://www.forbes.com/sites/rebekahbastian/2021/10/24/the-diversity-equity-and-inclusion-potential-of-nfts/}

\smallskip \noindent
Belk, R., Humayun, M., and Brouard, M. 2022. Money, possessions, and ownership in the Metaverse: NFTs, cryptocurrencies, Web3 and Wild Markets. \textit{Journal of Business Research} 153: 198--205.

\smallskip \noindent
Blodgett, S. L., Barocas, S., Daum\'{e} III, H., and Wallach, H. 2020. Language (technology) is power: A critical survey of ``Bias'' in NLP. arXiv preprint arXiv:2005.14050.

\smallskip \noindent
\c{C}a\u{g}layan Aksoy, P., and \"{O}zkan \"{U}ner, Z. 2021. NFTs and copyright: challenges and opportunities. \textit{Journal of Intellectual Property Law and Practice} 16(10): 1115--1126.

\smallskip \noindent
Chohan, R., and Paschen, J. 2023. NFT marketing: How marketers can use nonfungible tokens in their campaigns. \textit{Business Horizons} 66(1): 43--50.

\smallskip \noindent
Constantino, T. 2022. NFTs Get Religion as Vatican Creates a Digital Gallery. \url{https://www.fool.com/the-ascent/cryptocurrency/articles/nfts-get-religion-as-vatican-creates-a-digital-gallery/}

\smallskip \noindent
Crypto Altruism. 2023. Three Indigenous-led NFT Projects elevating Indigenous art and culture. \url{https://www.cryptoaltruism.org/blog/three-nft-projects-elevating-indigenous-art-and-culture}

\smallskip \noindent
Dean, T. 2022. Hype and hypocrisy: The high ethical cost of NFTs. \url{https://ethics.org.au/hype-and-hypocrisy-the-high-ethical-cost-of-nfts/}

\smallskip \noindent
Dignum, V. 2022. Relational artificial intelligence. arXiv preprint arXiv:2202.07446.

\smallskip \noindent
Flick, C. 2022. A critical professional ethical analysis of Non-Fungible Tokens (NFTs). \textit{Journal of Responsible Technology} 12(100054): 1--16.

\smallskip \noindent
Fowler, A., and Pirker, J. 2021. Tokenfication --- The potential of non-fungible tokens (NFT) for game development. In \textit{Extended Abstracts of the 2021 Annual Symposium on Computer-Human Interaction in Play,} 152--157.

\smallskip \noindent
Fraser, N., and Honneth, A. 2003. \textit{Redistribution or Recognition? A Political-Philosophy Exchange.} Verso.

\smallskip \noindent
Gibson, J. 2021. The thousand-and-second tale of NFTs, as foretold by Edgar Allan Poe. \textit{Queen Mary Journal of Intellectual Property} 11(3): 249--269.

\smallskip \noindent
Habermas, J. 1990. \textit{Moral Consciousness and Communicative Action.} London: Polity Press.

\smallskip \noindent
Hategan, V. 2024. Dead NFTs: The Evolving Landscape of the NFT Market. \url{https://dappgambl.com/nfts/dead-nfts/}

\smallskip \noindent
Huggard, E., and S\"{a}rm\"{a}kari, N. 2023. How digital-only fashion brands are creating more participatory models of fashion co-design. \textit{Fashion, Style and Popular Culture} 10(4): 583--600.

\smallskip \noindent
Ingold, L. 2022. NFTs: Risks, Rewards, Ethics. \url{https://sevenpillarsinstitute.org/nfts-risks-rewards-ethics/}

\smallskip \noindent
Kaczynski, S., and Kominers, S. D. 2021. How NFTs create value. \textit{Harvard Business Review.} \url{https://hbr.org/2021/11/how-nfts-create-value}

\smallskip \noindent
Kumar, N. 2025. 49 NFT Statistics 2025 --- Worldwide Data and Market Forecast. \url{https://www.demandsage.com/nft-statistics/}

\smallskip \noindent
Metablock. 2024a. NFTs and Historical Artifacts. \url{https://blog.metablock.co/nfts-and-historical-artifacts-073de79caa19}

\smallskip \noindent
Minaev, A. 2023. How Many NFTs Are There in 2024? \url{https://cryptodose.net/learn/how-many-nfts-are-there/}

\smallskip \noindent
Musamih, A., Salah, K., Jayaraman, R., Yaqoob, I., Puthal, D., and Ellahham, S. 2022. NFTs in healthcare: vision, opportunities, and challenges. \textit{IEEE Consumer Electronics Magazine} 12(4): 21--32.

\smallskip \noindent
Nguyen, J. K. 2022. Racial discrimination in non-fungible token (NFT) prices? CryptoPunk sales and skin tone. \textit{Economics Letters} 218: 1--4.

\smallskip \noindent
Okonkwo, I. E. 2021. NFT, copyright and intellectual property commercialization. \textit{International Journal of Law and Information Technology} 29(4): 296--304.

\smallskip \noindent
Sandel, M. J. 2020. \textit{The Tyranny of Merit.} Penguin Random House.

\smallskip \noindent
Sen, A. 2001. \textit{Development as Freedom.} Oxford University Press.

\smallskip \noindent
Smith, S. V. 2023. Is it RIP for the NFT? \url{https://www.marketplace.org/story/2023/12/22/nft-market-crashes-value-worthless}

\smallskip \noindent
Taherdoost, H. 2022. Non-fungible tokens (NFT): A systematic review. \textit{Information} 14(1): 1--12.

\smallskip \noindent
Watkins, I. 2021. Non-fungible tokens (NFT) and the evolution of art. In \textit{The Youth of the 21st Century: Education, Science, Innovations.} \url{https://rep.vsu.by/bitstream/123456789/30743/1/278-280.pdf}

\smallskip \noindent
Web3Sense. 2025. Web3Sense. \url{https://www.web3sense.ai/}

\smallskip \noindent
Young, I. M. 2020. Justice and the Politics of Difference. In \textit{The New Social Theory Reader,} 261--269. Routledge.

\smallskip \noindent
Zhang, Y., Chen, Z., Xu, C., Zhang, L., and Tong, X. 2022. Are racial and gender inequalities emerging in the NFT artwork? A visual exploration of CryptoPunks. 

\end{document}